\documentclass[10pt]{iopart}
\begin{document}
\title{Comments on the Invariance of Physical Laws Under Particle Re-Arrangement}
\author{M.\ Spaans}
\address{Kapteyn Astronomical Institute, University of Groningen; spaans@astro.rug.nl}
\begin{abstract}
Observationally and experimentally, the laws of physics express how particles interact.
Conversely, physical laws should be invariant under any re-arrangement of those particles, e.g., the laws of gravity
do not change if one re-arranges the stars in the sky.
To explore the physical meaning of these assertions, arguments are presented that show how the freedom
of particle re-arrangement leads to an identical twin associated with any photon, i.e., nature sees double.
These twins can become spatially separated for astronomically distant objects and are special in that absorption of
the one causes the disappearance of the other.
A tilting detector then leads to brightness variations across an image for twin separations on the order of the
detector size.
\end{abstract}
%\pacs{12.10.-g, 98.80.Bp, 04.20.Gz}
%\submitto{\JPA}
%Comment out if separate title page not required
%\maketitle

\section{Invariance of Physical Laws Under Particle Re-Arrangement}

Operationally, one can view any experiment as a sequence of interactions between particles.
From the symmetries and systematics in the outcomes of these interactions one can then derive physical laws.
As such, the laws of physics are expressed by the interactions between particles. However, one can equally well
assert that all the different sorts of interactions between particles constitute physical laws since interactions
are the only means by which observers can derive physical laws.

E.g., suppose that a proton and an electron interact electromagnetically through the exchange of photons
and that one derives Maxwell's equations from the outcomes of experiments.
Imagine then that one independently moves the proton and electron around.
This causes interactions since forces must act under particle re-arrangement, but Maxwell's equations remain
valid because one cannot change those by re-arranging particles. Afterall, what one does is to change the
initial and/or boundary conditions of some experiment. These conditions are needed to solve the equations of motion,
but do not alter the equations.
Similarly, re-arranging the stars in the sky does not change the laws of gravity either.

Since physical laws describe {\it how} something happens rather then {\it what} happens,
consider the unusual position that 1) the combined interactions between all particles in the universe constitute physical
laws throughout space-time, that 2) all particles involved in these interactions can be re-arranged without altering the
laws, and that 3) all re-arrangements in turn provide the {\it only} limitations on how these interactions occur.
Thus, 3) takes freedom of re-arrangement alone to constrain interactions, i.e., the physical presence of particles necessary
for interactions {\it follows} just from 3).
In this way nature sees only how particles interact through re-arrangements, {\it precluding} any background dependent
distinction between an experiment and its observer\cite{1,2}.
These properties of physical laws are denoted general invariance (GI). 
\hfill\vfill\eject

\section{Observational Effects}

To test GI, consider a particle A that emits a photon ``$\rightarrow$''.
GI then demands that particles in the universe can be re-arranged, so with the photon in flight, without altering physical laws.
One is then confronted with two mutually exclusive interactions for how to detect the photon and to incorporate it into
physical laws: A$\rightarrow$B and A$\rightarrow$A, for an emitter A that is its own absorber, hiding the photon from all
other observers, and an arbitrary particle B$\ne$A representing all other observers.
These are the only two possibilities since re-arrangement renders all B$\ne$A particles equivalent, i.e., all can placed on the photon path,
while re-arrangement cannot turn the particle that has emitted the photon into one that has not or vice versa.
Interactions A$\rightarrow$B and A$\rightarrow$A need not be equally probable, or even likely, but must always be physically possible
under GI. This then requires an identical twin photon (same energy and wave vector) because a photon in flight is simultaneously on its
way to an interaction with A and to an interaction with the rest of the universe. I.e., nature is forced to {\it see double} since it only
has re-arrangements as a constraint on interactions.
The probability for interaction of each photon is ${{1}\over{2}}$ because one can choose either irrespective of which
interaction occurs. Re-arranging ``$\rightarrow$'' to travel in the opposite direction leads to identical conclusions
since two photons are again needed a priori in order to allow one to travel to B and one back to A.
Obviously, if either photon is destroyed (or reflected) then both must disappear (or redirect) since a pair
derives from re-arrangement applied to an existing photon.
So the expectation value of a photon pair property $P$ (like angular momentum) is an average, $<P>={{1}\over{2}}P_1+{{1}\over{2}}P_2$
with $P_1=P_2$, and no conservation laws are violated when the twin state (1,2) is induced by GI.

It may be possible to detect this twinning since quantum mechanics forbids the two photons, say of wavelength $\lambda$,
to have wave vectors that are aligned to better than a precision $\sim l_P/\lambda$, for the Planck length $l_P$.
Imagine that the photon twins travel across the universe without any interaction.
A physical separation $\delta =d(l_P/\lambda )$ occurs between the photon twins
when their traversed distance is $d$. E.g., $\delta =10$ cm for 10 keV photon twins from a source
$10^{26}$ cm ($10^8$ lightyears) away.
Conversely, setting $\delta\approx\lambda$ yields a distance $d_s\approx\lambda^2/l_P$ for which the separation
is $\approx\lambda$, large enough to render the twins mutually independent.
Photon twins will not be separated much in time since $\delta /d$ is very small.
So absorption in a plane perpendicular to the source line-of-sight randomly picks out one of the twins,
yielding a uniform signal.
Subsequent tilting of the detector then favors detection of twin members with the shortest travel distance to the source, and
causes brightness variations for $\delta\approx L$ and a detector size $L$.
Twins should also exist for massive vector bosons and gluons, but these do not travel far. General relativity is purely geometric
and can be naturally expressed by the differences between inertial frames that are induced by particle re-arrangement.

If twin photons are detected, then one can speculate on how particle re-arrangement takes place physically.
Macroscopically (e.g., the motion of light and matter in curved space) this is well understood,
but at the Planck scale there are more exotic possibilities, e.g.,
microscopic wormholes through which particles travel\cite{3} or the quantum uncertainty in particle space-time labels. Topology is
best suited to express these non-local re-arrangements and one should look for a topology that allows any particle, also
the photon twins, to choose between an A$\rightarrow$B and A$\rightarrow$A path\cite{4}.


\begin{thebibliography}{}

\bibitem{1}
L.\ Smolin, arXiv:hep-th/0507235

\bibitem{2}
G.\ 't Hooft, arXiv:hep-th/0707.4568

\bibitem{3}
J.A.\ Wheeler, Phys.\ Rev.\ 97, 511 (1959)

\bibitem{4}
M.\ Spaans, arXiv:gr-qc/0705.3902; Nuc.\ Phys.\ B 492, 526 (1997)

\end{thebibliography}
\end{document}